\documentstyle[aas2pp4]{article}
\def\xxi0{\hbox{${\xi^\prime}_0$}}
\def\th0{\hbox{$\theta_0$}}
\def\r0{\hbox{$\varpi_0$}}

\def\f0{\hbox{$\psi_0$}}

\def\v1{\hbox{$v_{K1}$}}
\def\ro1{\hbox{$\rho_1$}}
\def\n1{\hbox{$n_1$}}
\def\b1{\hbox{$B_1$}}

\def\msol{\hbox{${\rm M}_\odot $}}

\def\lsol{\hbox{${\rm L}_\odot $}}
\def\kelvin{{\rm K}}

\def\kgauss{{\rm kG}}

\def\angstrom{\hbox{\AA}}

\def\myr{\mbox{\rm Myr}}

\def\refeqm#1{(\ref{eq:#1})}
\def\refeqp#1{[\ref{eq:#1}]}

\def\subsectn#1{
    \addtocounter{subsection}{1}
    \setcounter{subsubsection}{0}
    {\ifnum\value{subsection}>1 \vskip 0.25 truein \penalty 10000 \fi}
    \penalty 10000
    {\centerline {\it \thesubsection.\ #1}}
    \penalty 10000      
    \bigskip
    \penalty 10000
    \index{#1}
    }
\def\subsubsectn#1{
    \addtocounter{subsubsection}{1}
    {\ifnum\value{subsubsection}>1 \vskip 0.125 truein \penalty 10000 \fi}
    \penalty 10000
    {\centerline {\it \thesubsubsection.\ #1}}
    \penalty 10000
    \bigskip
    \penalty 10000
    \index{#1}
    }
\def\beq{\mbox{$B_{\rm eq}$}}
\def\fb{\mbox{$fB$}}
\def\teff{\mbox{$T_{\rm eff}$}}

\begin{document}
\thispagestyle{empty}
\title{A Physical Limit to the Magnetic Fields of T Tauri Stars}
\author{Pedro N. Safier}
\affil{University of Maryland,\\
 Laboratory for Millimeter-Wave Astronomy,\\
 College Park, MD 20742;\\ e-mail: {\it
    safier@astro.umd.edu\/}}
\begin{abstract}
Recent estimates of magnetic field strengths in T Tauri stars
yield values $B=1$--$4\,{\rm kG}$. In this paper, I present an upper
limit to the photospheric values of $B$ by computing the equipartition
values for different surface gravities and effective temperatures.
The values of $B$ derived from the observations exceed this
limit, and I examine the possible causes for this discrepancy.
\end{abstract}
\vspace{1.0in}
\centerline{\large To appear in {\it The Astrophysical Journal Letters}}
\setcounter{page}{0}         %
%
%
\section{Introduction}
Magnetic fields are believed to play a fundamental role in the
structure and evolution of T Tauri stars. Unfortunately, their
detection and measurement in these objects is a very difficult task. 

Early attempts to detect
magnetic fields in T Tauri stars through Zeeman polarization were
unsuccessful (Johnstone \& Penston 1986\markcite{jp86},
1987\markcite{jp87}), and 
only recently such detections are forthcoming through the measurement
of Zeeman broadening of photospheric lines in a handful of objects (
Basri et al. 1992\markcite{bmv92}; Guenther
1997\markcite{gue97}). These measurements yield disk-averaged, photospheric
field strengths of order a few kG.

These observations do not provide a direct measurement of $B$; instead,
what is measured is the difference in broadening of
magnetically-sensitive and 
insensitive lines (Robinson 1980\markcite{rob80}; Marcy
1982\markcite{mar82}; Saar 1988\markcite{saar88}).
To find the value of $B$, one has to fit the data
with a stellar atmosphere model  and  vary $B$ and
$f$---the fraction of stellar surface that is magnetized---until a
good match is found. The result of this matching
procedure is the product \fb, and independent determination of $B$
and $f$ is much more problematic and sensitive to modeling assumptions
(see, e.g., Solanki 1992\markcite{sol92} and references
therein). Furthermore, when equivalent widths are used in lieu of
measuring the  line broadening, the deduced \fb\ is also very
sensitive to the input model atmosphere (Basri et al. 1992; Basri \&
Marcy 1994\markcite{bama94}). 

The purpose of this paper is to present 
a physical constraint on the maximum value of \fb; this upper limit on
the magnetic field strength  is important to estimate $B$ from the
\fb-values obtained from the data, and may help to uncover potential
errors in the modeling assumptions. 

The basic physics is rather
well known, and has been successfuly applied to the study of the
Sun. Magnetic fields 
are excluded from closed 
circulation patterns when the field lines are perpendicular to the
axis of rotation of the fluid (see, e.g., Parker 1979\markcite{enpbook79}, 
ch. 16 and references therein); as a result of this general principle
the solar convection sweeps  the 
field into the downdraft regions of the eddies. Convective
heat transport is strongly inhibited within the field (Biermann
1941\markcite{bier41}), and, through the so-called superadiabatic
effect (Parker 1978\markcite{enp78}), the downflowing gas further
cools and evacuates the
magnetized gas; as a result, the external gas pressure is larger than
that inside the tube, and the flux tube shrinks until the total
internal pressure (thermal plus magnetic)  balances the external pressure.
Thus, the solar magnetic field is broken up and compressed into highly
evacuated flux tubes.

Although we cannot resolve features on
the surface of other stars, it is a reasonable inference that, in
complete analogy with the Sun, the magnetic field in stars
with outer convection zones---or convective throughout, as in the case
of T Tauri
stars---is concentrated into discrete flux tubes,
which are in pressure equilibrium with the unmagnetized gas. If this
is true, then an {\it upper\/} limit on $B$ can be derived if the pressure of
the unmagnetized 
gas is known and the gas pressure inside the flux tubes is
ignored\footnote{Note that this argument does not 
  apply to stars with outer radiative zones, where there are no known
  mechanisms to isolate and concentrate the field. The Ap stars are a prime
  example, with measured fields as high as $30\,\kgauss$ in the case of
  HD 215441 (Babcock 1960\markcite{bab60}), a value much larger than
  that obtained by assuming 
  pressure balance. Therefore, these fields are likely to thread
  the entire surface, and probably are fossil, rather than
  dynamo-generated, fields. In the case of T Tauri stars, after the
  onset of convection the fossil field is quickly destroyed by the
  resulting turbulent difussivity. K\"{u}ker \& R\"{u}diger
  (1997\markcite{kuk97}) estimate that in a typical TTS the fossil
  field is destroyed in $\sim300\,{\rm yr}$ after the onset of
  convection. Therefore, the magnetic field in TTS must be
  dynamo-generated.}. These arguments have been 
extensively applied to active dwarfs
(see, e.g., Saar 1996\markcite{saar96}),
but, so far, they have never been used to study the magnetic fields
in T Tauri stars. 

In \S 2 I derive an expression for the maximum equipartition value of $B$
as a function of the stellar and atmospheric parameters, and in \S 3 I
compare the results from \S 2 with current observational estimates. My
conclusions follow in \S 4.

\section{Equipartition Fields}

 Consider a straight flux tube that is
initially perpendicular to the photospheric surface at some fiducial
height. The condition of 
lateral pressure equilibrium, i.e., the flux tube neither expands nor
contracts, is given by 
\begin{equation}
  \label{eq:1}
  \frac{B^2}{8\pi} = \Delta P,
\end{equation}
where $\Delta P$ is the difference between the external and internal
non-magnetic pressures, $P_i$ and $P_e$, respectively, and the external
medium is unmagnetized. At any height, an upper limit to $B$ is given
by the equipartition field $\propto (P_e)^{1/2}$, and because
temperature and pressure decrease with height in the photosphere, the
maximum detectable field is the equipartition field at the optical
depth where the continuum is formed ($\tau=2/3$),
\begin{equation}
  \label{eq:2}
  \beq = \left[8\pi\,P_e(\tau=2/3)\right]^{1/2}.
\end{equation}

In principle, $P_e$ includes the contributions from both  thermal pressure
and ram pressure due to convective motions ($P_T$ and
$P_v$, respectively) because in a convective atmosphere
the magnetic field outside dark spots is concentrated  where the
convective downdrafts 
are located (see, 
e.g., Stein et al., 1992\markcite{stal92}). However, it is easy to
show that $P_T \gg 
P_v$, and the argument is the following. 

The ratio $P_v/P_T$ can
be written as
\begin{equation}
  \label{eq:pvpt1}
  \frac{P_v}{P_T} = \frac{1}{3}\, \gamma\,
  \left(\frac{\bar{v}}{v_s}\right)^2,
\end{equation}
where $\gamma$ is the ratio of the specific heat capacities, $\bar{v}$ is the
average speed of a (turbulent) convective element, and $v_s$ is the
adiabatic speed of sound. 
Although recent
simulations of compressible convection (Cattaneo \& Malagoli,
1992\markcite{cm92}) show  that {\it horizontal\/} surface flows can be
intermittently transonic, the downflows are subsonic, and, because
$\gamma \le 5/3$, it follows that  $P_v/P_T\le 5/9$. A more
stringent limit on this ratio is given by estimating $\bar{v}/v_s$
using the mixing-length theory of convection, which gives (see, e.g.,
Cox \& Giuli 1968\markcite{cg68}, eq. 14.64)
\begin{equation}
  \label{ptpv2}
  \frac{\bar{v}}{v_s} = 0.4\, \left(\frac{L_\ast}{1.4\,\lsol}\right)^{1/3}\,
                              \left(\frac{\teff}{4000\,\kelvin}\right)^{-1/2}\,
                              \left(\frac{\mu}{1.7}\right)^{1/2},
\end{equation}
where $L_\ast$, $\teff$, and $\mu$ are, respectively, the stellar
luminosity, effective temperature, and photospheric mean molecular
weight. Therefore, for the stellar parameters of a typical TTS,
$P_v/P_T\approx 0.2$, and 
by neglecting $P_v$ the error introduced in \beq\ is $\sim
10$\%---much smaller than the observational errors (see
below). Henceforth I will neglect the contribution of ram
pressure\footnote{The same result, essentialy, applies to the Sun,
  where the Maxwell stresses at the surface are an order of magnitude
  larger than the Reynolds stresses.} to \beq, and use $P_e = P_T$.

The value of  $P_e(\tau=2/3)$ follows from the condition of
hydrostatic equilibrium, and to a very good degree of approximation
(generally, better than a 
factor of $2$ or $3$; see, e.g., Cox \& Giuli 1968\markcite{cg68}),
one can write 
\begin{equation}
  \label{eq:3}
  P_e(\tau=2/3) = \frac{2}{3}\,\frac{g}{\kappa},
\end{equation}
where $\kappa$ is the Rosseland mean opacity.
The gas temperature $T$ at $\tau=2/3$ is assumed to be the effective
temperature of the star, \teff, and, because $\kappa$ is a function of
density $\rho$ and $T$, eq.~\refeqm{3} is an implicit equation for
$\rho(T=\teff)$ for a given \teff, $g$, and
$\kappa(\rho,\teff)$. Thus, one finally obtains 
\begin{equation}
  \label{eq:4}
  \beq = \left[\frac{16\pi}{3}\,\frac{g}{\kappa}\right]^{1/2}.
\end{equation}

I have computed \beq\ for different values of \teff\ and $g$ by
solving eq.~\refeqm{3} using the Alexander \& Ferguson (1994\markcite{af94})
Rosseland mean opacities, and using eq.~\refeqm{4}. To gauge the
magnitudes of the errors introduced by the approximation used to
derive eq.~\refeqm{3}, I have also computed \beq\ by obtaining
$P_e(T=\teff)$ from the detailed stellar-atmosphere models by Guenther
et. al (1992) and Allard \& Hauschildt (1995). The comparison between
the two sets of \beq\ so obtained is presented in Table~1.

\begin{deluxetable}{lccrl}
\tablewidth{33pc}
\tablecaption{Comparison Between $\beq$ from Equation (6) and Detailed Stellar Atmospheres}
\tablehead{
\colhead{Sp. Type} &  \colhead{$\log g$} & \colhead{$\beq\,\rm{(kG)}$\,\tablenotemark{a}} &
 \colhead{$\,\Delta\,$\tablenotemark{b}} & \colhead{Atmospheric Model}}
\startdata
Sun    &  $4.4$ & $1.2$ & $0.08$   & Guenther et al. 1992 \nl
K5     &  $4.5$ & $1.8$ & $0.06$   & Allard \& Hauschildt 1995 \nl
M1     &  $4.0$ & $1.5$ & $-0.13$  & Allard \& Hauschildt 1995 \nl
M5     &  $4.0$ & $1.5$ & $-0.36$  & Allard \& Hauschildt 1995 \nl
M5     &  $5.0$ & $4.8$ & $-0.52$  & Allard \& Hauschildt 1995 
\enddata
\tablenotetext{a}{Computed from eq.~(6).}
\tablenotetext{b}{Fractional difference; the equipartition field
  computed from a detailed atmospheric model is given by $\beq\,[1+\Delta]$.}
\end{deluxetable}

The results in Table~1 show that for spectral type K5, or earlier, the
approximation used in eq.~\refeqm{3} results in an underestimate of
the field by less than $10$\%. On the other hand, for later spectral
types the field is {\it overestimated} by, at most, a factor of $2$,
but this error decreases with decreasing gravity. The reason for this
larger discrepancy at cooler temperatures is the increase in opacity
due to H$_2$O  once the gas temperature drops below
$3500\,\kelvin$ (Alexander \& Ferguson 1994), which results in a
strong non-monotonic behavior of 
$\kappa$ with optical depth in the outer layers.

Therefore, for the typical gravities of T Tauri stars ($\log\,g\la
4.0$) and the spectral types currently accessible to
Zeeman-broadening measurements (spectral type K7 or earlier),  the
equipartition fields predicted from eq.~\refeqm{4} are too small by,
at most, $10$\% when compared with detailed atmospheric models.. These
errors are much smaller than the error bars 
for the measured $fB$'s (see Figure 1 below).

It is important to keep in mind that the {\it measured\/} field depends on the
{\it geometrical\/} depth where the emission originates. Because
flux tubes are strongly evacuated, the optical depth inside a
magnetized region is smaller than in the surrounding atmosphere.
If the flux tube is slender enough, it will be mostly transparent to
the radiation crossing its walls, 
and, because of its lower density, it may
appear as a bright feature against the continuum from its surroundings,
as the network fields in the Sun do. In the solar case, the enhanced
temperature compensates for the lower optical depth, and the values of $B$
measured in these bright points compare well with
$\beq(T=\teff)$. If the Sun is any guidance, then $\beq(T=\teff)$
should be also a good approximation to the fields outside spots  in other
stars.

On the other hand, larger
features like 
sunspots are much more evacuated and cooler than slender flux tubes,
and at a fixed geometrical height the optical depth is much smaller
inside the spot. Thus, one can see deeper
into a sunspot, and they indeed appear as dimples on the Sun's surface
when seen near the limb---a phenomenon known as the Wilson
depression. Because the continuum from  a spot originates at a larger
geometrical depth, the external pressure there is larger, and, therefore, the
fields measured inside sunspots are larger than those from slender
features like network fields. In the Sun, the Wilson depression can
amount to a few scale heights, and thus the measured fields in
sunspots are $\sim 2 \beq(T=\teff)$, i.e., of order $3\,\kgauss$
(although, sometimes, values as high as $5\,\kgauss$ have been
measured). In complete analogy with the Sun, one should not expect
$\beq(T=\teff)$ to be a useful limit---within a factor of a few---to
the fields in stellar dark spots. The equipartition argument still
applies; however, to derive a useful limit on $B$ it is necessary to
know the detailed thermal structure of the spot, a problem that still
lacks a  full solution even for the Sun.

\section{Comparison with Current Measurements}
At present, only two set of magnetic-field measurements for T Tauri
stars are available\footnote{Donati et al. (1997)\markcite{donal97}
  detected circular polarization in V410 Tau and HDE 283572
  (V987~Tau), but they did not derive a value of $B$ from their 
  measurements. Guenther \& Emerson (1996\markcite{gue96}) attempted
  to measure the fields in Tap 35 and V410 Tau to demonstrate the
  potential of infrared lines in Zeeman studies, but they obtained
  upper limits to $B$ that are consistent with the measurements of
  Basri et al. (1992).}: the measurements by Basri et al. (1992) and 
Guenther (1997). These observations 
measure the broadening of Fe\,I lines in the 
wavelength range $5000$--$7000\,$ \angstrom ; at these wavelengths,
a dark spot with temperature $\teff - \Delta T$ that covers a fraction
$f_s$ of the surface of a star with
$\teff$ contributes a fraction $f_s\,B_\lambda(\teff - \Delta T)/(1-f_s)
B_\lambda(\teff)$ of the total emission, where $B_\lambda$ is the
Planck function. Using the typical values 
$f_s=0.3$, $\Delta T = 1000\,\kelvin$, and $\teff=4000\,\kelvin$ (see,
e.g., Bouvier et al. 1993\markcite{bal93}), one finds that starspots
contribute less than $7$\% to the light in the range $5000$--$7000\,$
\angstrom .
Therefore, current measurements
are insensitive to the magnetic fields in the center of dark spots,
where $B$ can be larger than \beq\ at $T=\teff$.

Figure 1 is a comparison between the {\it lower\/} limits to $B$ derived
from the observations (because $f\le 1$) and \beq\ calculated from
\refeqm{4} as a function of \teff\ and different surface gravities.
Also shown is the value of \beq\ for the Sun; comparison of this value
with the one derived from \refeqm{4} for a G2 V star gives a measure
of the typical errors (Table 1) in the constant-$g$ loci in  Figure 1  for
the spectral types currently accessible to
Zeeman-broadening measurements (spectral type K7 or earlier). Note
that these errors are much smaller than the observational error bars.

\begin{figure}[h]
\epsscale{0.1}
\label{fig1}
\plotfiddle{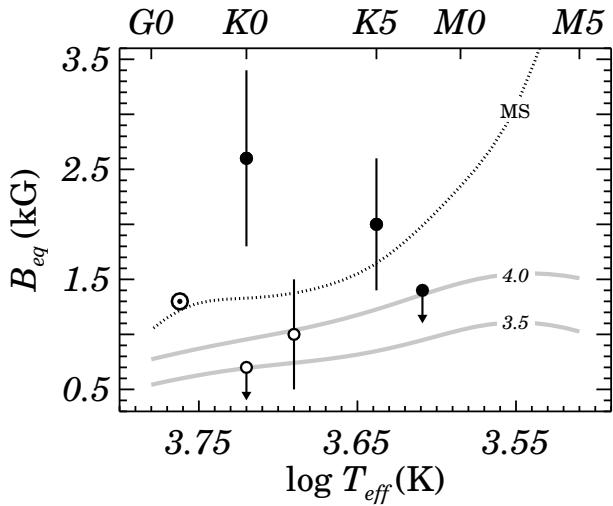}{3.0in}{90.}{40}{40}{150}{0}
\caption{The equipartition magnetic field, \beq, as a function of
effective temperature \teff\ and different values of surface gravity
$g$. Loci of constant $g$ are shown as {\it grey\/} lines and are
labeled by the value of $\log g$. The field strengths derived by
Guenther (1997)  are shown as {\it filled\/} circles (from
{\it left\/} to {\it right\/}): T Tau, Lk Ca 15, and Lk Ca 16; the
values of $B$ from the observations by Basri et al. are shown as {\it
  white\/} circles (from {\it left\/} to {\it right\/}): TAP 10 and
TAP 35. The {\it arrows\/} indicate upper limits.
Also shown is the locus of $\beq(\teff)$ for the main-sequence ({\it
  dotted\/} line) and the value of \beq\ for the Sun from the Guenther
et al. (1992) model is indicated by $\odot$.}
\end{figure}

There is a troubling trend in Figure 1, which is most obvious for the
Guenther (1997) data. The field-strengths derived by this author are above the
main-sequence values of \beq. In other words, the only way for these
objects to have magnetic fields  as strong as implied by the data is for them
to be {\it more\/} compact ($\log g \ga 4.5$) than a main-sequence
star of the same 
spectral type. Obviously, this is impossible; moreover,
$\log g \la 3.5$ for classical TTS such as those in the Guenther
(1997) dataset, and, also,  the
absorption lines form at $\tau < 2/3$.

Furthermore, since I have assumed $f=1$ in
deriving a value of $B$ from the measured $fB$,  these values, again,
are {\it lower limits\/} to the field implied by the
observations. Therefore, the measurements of Guenther (1997)
seem to overestimate $B$ by at least a factor of
$>2\,(\frac{2/3}{\tau})^{1/2}$ in the case of 
Lk~Ca~15 and a factor $>2.5\,(\frac{2/3}{\tau})^{1/2}$ for T Tau,
respectively.  

The measurements of Basri et al. (1992) for TAP 35 seem compatible
with the expected \beq, in particular because they assumed $\log g=4.0$
in their models. However, if $f<0.65$ their measurements would imply
a value of $B$ larger than the equipartition field in a main-sequence
star of the same spectral type. Moreover, they found that if
$\log g=3.5$, instead of $\log g=4.0$, then a stronger field is required to
match the data; this is
inconsistent with the dependence of \beq\ on $g$ at fixed
\teff\ (Figure 1). Therefore, there are also problems with
the Basri et al. (1992) measurements, and the implied fields are
larger than \beq.
\begin{figure}[h]
\epsscale{0.1}
\label{fig2}
\plotfiddle{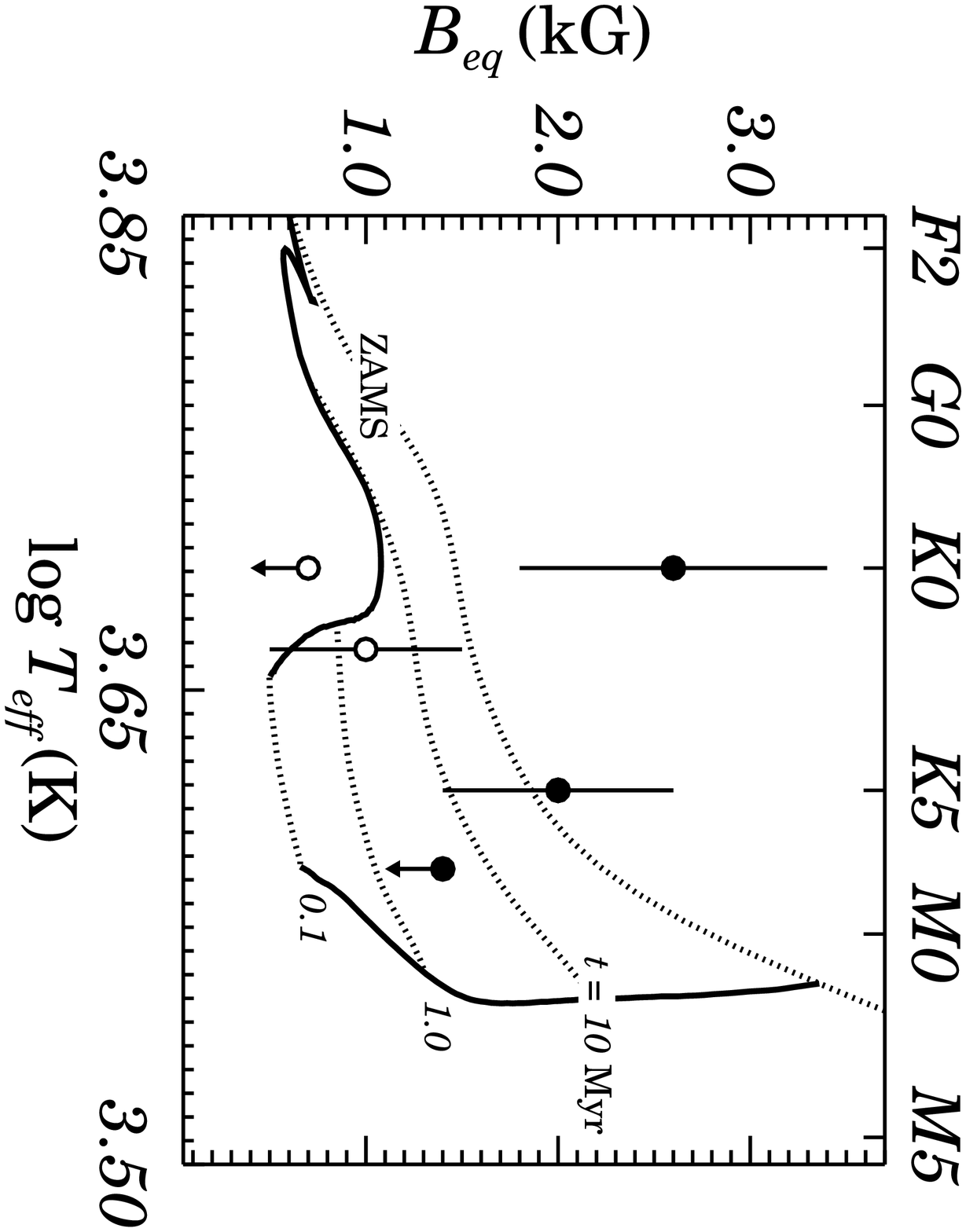}{3.0in}{90.}{40}{40}{145}{0}
\caption{The equipartition magnetic field, \beq, as a function of
effective temperature \teff\ along the pre-main-sequence isochrones of
D'Antona \& Mazzitelli (1994). The loci of constant age $t$ are shown
as {\it dotted\/} lines and are labeled by the value of $t$. The {\it
  solid\/} lines indicate the paths in the $(\teff,\beq)$-plane for 
a $1.5\,\msol$ ({\it left\/}) and a $0.4\,\msol$ star ({\it right\/}) as
they evolve towards the zero-age main-sequence. The other symbols have the
same meaning as in Figure~1.}    
\end{figure}

Another perspective on this inconsistency is presented in Figure
2. This figure is a plot of \beq\ as a function of \teff\ along the
pre-main-squence isochrones of D'Antona \& Mazzitelli (1994); also
shown are the paths in the $(\teff,\beq)$-plane traced by two stars with
$M=1.5\,\msol$ and $M=0.4\,\msol$, respectively, as they evolve
towards the zero-age main-sequence. 
These results show that, for a given value of the stellar mass, \beq\
increases with age; therefore, if one takes $\approx 10\,\myr$ as the
age limit for the T Tauri phase, the results in Figure 2 imply that in
TTS of spectral type K7 or earlier $B\la 1.5\,\kgauss$ outside dark spots.

\section{Conclusions}

The equipartition field \beq\ at any level in a stellar atmosphere is an
upper limit to the magnetic field strength in a flux tube in pressure
equilibrium with the surrounding, non-magnetized gas. I have  computed
\beq\ for the physical regime appropriate to T Tauri stars, and shown
that the field strengths derived from current Zeemann-broadening
measurements exceed this upper limit by large factors, and therefore
are unphysical.

What are the causes for this discrepancy?

Because the  derivation of $B$ from the measured line widths requires
detailed modeling of the emergent line profiles, the fact that current
models yield $B>\beq$ indicates that some piece of physics is
missing. To be fair to the Zeeman-broadening technique, when the
Zeeman components are resolved this method can be very precise
(see, e.g., the recent measurements of $B$ in $\epsilon$~Eri by
Valenti et al., 1995\markcite{val95}), whereas Basri et al. (1992) based
their analysis on the equivalent widths of the lines (due to the
faintness of TTS and the high signal-to-noise required by this
method), and Guenther 
(1997) used an autocorrelation analysis without benefit of a detailed
atmospheric model. However, it is hard to imagine how the use of
equivalent widths rather than line profiles can explain the finding by
Basri et al. (1992) that a lower surface gravity requires a larger
magnetic field to match the data, when the field is supposed to
decrease with decreasing gravity (see eq.~\refeqp{4}). Therefore, the
source of the discrepancy is, most likely, the input physics of the
models.

In particular, current models for TTS use the same atmospheric
structure for the magnetic and non-magnetic parts of the photosphere,
whereas at any given height the gas pressure in a flux tube has to be
lower than that of its surroundings for pressure equilibrium to
obtain. Because the measured field depends on the geometrical height
where the lines are formed, and the flux tubes must be highly
evacuated, using the same atmospheric structure for
the magnetized and quiet regions of the photosphere is not correct,
and is the most likely origin of the discrepancy.
Therefore, this and other refinements (see Landolfi et
al. 1989\markcite{lll89})  must be
incorporated to derive a more accurate estimate of $B$ in TTS.
In addition, Valenti et al. (1995) showed that the Zeeman-broadening
method is much more immune to the details of the model atmosphere when
infrared lines are used. Infrared measurements  should be pursued
also, and one hopes that 
the discrepancy with the results presented here will disappear.

I wish to thank Robert Rosner for a conversation during which the idea
for this paper was conceived, and Lee Mundy, Steve
Stahler, and Stephen White for their careful reading of the manuscript. 
 I am also indebted to an anonymous referee, who provided comments
 and criticisms that significantly improved an earlier version of this paper.
This work was  supported by NSF grant AST9613716 to the Laboratory
for Millimeter-Wave Astronomy at the University of Maryland.

\end{document}